\documentclass[prd,letterpaper,twocolumn,preprintnumbers,nofootinbib]{revtex4}

\usepackage{amsmath,amssymb}
\usepackage{graphicx}
\usepackage{footmisc}
\usepackage[hyperfootnotes=false,colorlinks,citecolor=blue]{hyperref}

\def \L {\mathcal{L}} 

\def \epsilon {\varepsilon} 

\def \vec#1{{\boldsymbol{#1}}}
\newcommand{\dd}{{\rm d}}
\newcommand{\ii}{{\rm i}}

\allowdisplaybreaks

\begin{document}

\title{Q-balls in polynomial potentials}

\author{Julian Heeck}
\email[Email: ]{heeck@virginia.edu}
\thanks{ORCID: \href{https://orcid.org/0000-0003-2653-5962}{0000-0003-2653-5962}.}
\affiliation{Department of Physics, University of Virginia,
Charlottesville, Virginia 22904-4714, USA}

\author{Mikheil Sokhashvili}
\email[Email: ]{ms2guc@virginia.edu}
\thanks{ORCID: \href{https://orcid.org/0000-0003-0844-7563}{0000-0003-0844-7563}.}
\affiliation{Department of Physics, University of Virginia,
Charlottesville, Virginia 22904-4714, USA}

\hypersetup{
pdftitle={Q-balls in polynomial potentials},   
pdfauthor={Julian Heeck, Mikheil Sokhashvili}
}


\begin{abstract}
Bosons carrying a conserved charge can form stable bound states if their Lagrangian contains attractive self-interactions. Bound-state configurations with a large charge $Q$ can be described classically and are denoted as Q-balls, their properties encoded in a non-linear differential equation. Here, we study Q-balls in arbitrary polynomial single-scalar-field potentials both numerically and via various analytical approximations. We highlight some surprising universal features of Q-balls that barely depend on the details of the potential. The polynomial potentials studied here can be realized in renormalizable models involving additional heavy or light scalars, as we illustrate with several examples.
\end{abstract}

\maketitle



\section{Introduction}

Q-balls are interesting examples of large bound states,  in the simplest scenario consisting of $Q\gg 1$ complex scalars $\phi$ with conserved global charge $Q(\phi)=1$. Assuming an attractive force between these scalars, Q-balls form the lowest-energy configuration for a fixed charge $Q$ and are hence stable~\cite{Coleman:1985ki}. Due to the large amount of scalars residing in the Q-ball, it can be described classically as a spherically-symmetric solution to the non-linear Lagrange equations, also known as a non-topological soliton~\cite{Lee:1991ax}. As emphasized already by Coleman in his seminal paper on these objects~\cite{Coleman:1985ki}, a renormalizable \emph{quantum} field theory for $\phi$ by itself does not provide the required attractive interactions, but it is possible to construct multi-field models that lead to the required terms in the scalar potential~\cite{Friedberg:1976me,Coleman:1985ki,Lee:1991ax,Kusenko:1997zq,Nugaev:2019vru,Lennon:2021zzx}.

Even in effective single-field potentials it is typically impossible to analytically solve the underlying non-linear differential equations, save for some special and often unphysical examples~\cite{Rosen:1968mfz,Rosen:1969ay,Theodorakis:2000bz,MacKenzie:2001av,Arodz:2008jk,Gulamov:2013ema}. In general, we have to satisfy ourselves with numerical solutions or analytical approximations, which include Coleman's thin-wall approximation~\cite{Coleman:1985ki} (valid for very large Q-balls with thin surface region) and Kusenko's thick-wall approximation~\cite{Kusenko:1997ad} (valid for small but dilute Q-balls).
These approximations allow for an improved understanding of Q-balls that is difficult to obtain from numerical scans and in particular cover the regions of parameter space that are challenging to investigate numerically~\cite{PaccettiCorreia:2001wtt,Heeck:2020bau}.

The simplest consistent realization of Coleman's large Q-balls~\cite{Coleman:1985ki} requires a scalar potential for $\phi$ with a mass term $m_\phi^2|\phi|^2$, an attractive interaction term $\propto - |\phi|^p$, and a term that stabilizes the potential at large field values $\propto + |\phi|^q$, with $2<p<q$. We will study Q-balls in such potentials as a function of $p$ and $q$, mostly restricted to integer exponents. We provide analytical approximations, including the thin- and thick-wall limits, and compare them to numerical solutions. In the thin-wall, or large~$Q$, regime, we find remarkably simple analytical solutions for arbitrary $p$ and $q$.
Our model and notation is set up in Sec.~\ref{sec:model}. Sec.~\ref{sec:thin-wall} generalizes the thin-wall approximation of Ref.~\cite{Heeck:2020bau} to arbitrary $p$ and $q$, including the particularly easy-to-solve cases of equidistant exponents. 
Sec.~\ref{sec:thick} collects thick-wall results in our notation, restricted to the case $p=3$, as is this the only integer value for $p$ that gives stable Q-balls in the thick-wall regime.
In Sec.~\ref{sec:largen} we introduce a novel Q-ball approximation that is valid for $p,q\gg 2$ irrespective of the wall thickness.
In Sec.~\ref{sec:UV} we study some simple renormalizable multi-field models and discuss when and how they can be described by our effective polynomial potentials.
We discuss our results and conclude in Sec.~\ref{sec:conclusion}.
App.~\ref{app:rescaled_heavy} gives an alternative derivation of some results of Sec.~\ref{sec:UV}.

\section{Model}
\label{sec:model}

Using the mostly-minus Minkowski metric, we study single-field Q-balls~\cite{Coleman:1985ki} with a Lagrangian
\begin{align}
    \L = |\partial_\mu \phi|^2 - U(|\phi|)
\end{align}
 for the complex scalar $\phi$ that is invariant under a global $U(1)$ symmetry $\phi\to e^{\ii \alpha} \phi$, with constant $\alpha\in \mathbb{R}$, leading via Noether's theorem to a conserved charge $Q$, normalized here to $Q(\phi)=1$. $Q$ therefore counts the number of $\phi$ particles in a given field configuration.
The Euler--Lagrange equation takes the form
\begin{align}
    \partial_\mu \partial^\mu \phi +\frac{\partial U}{\partial \phi^*} =0\,,
    \label{eq:lagrange-equation}
\end{align}
for which we will discuss a particular set of solutions.
If the potential $U$ contains attractive interactions, a bound-state solution with charge $Q$ is possible that has the lowest energy among all configurations with the same charge, and is hence stable~\cite{Coleman:1985ki}. The potential needs to fulfill
\begin{align}
  \left.  \frac{\dd U}{\dd |\phi |}\right|_{\phi = 0} = 0\,, &&
  \left.  \frac{\dd^2 U}{\dd \phi\dd \phi^*}\right|_{\phi = 0} \equiv  m_\phi^2 >0\,, 
\end{align}
so that the vacuum $\phi =0$ is stable and the $U(1)$ unbroken, and $U(|\phi|)/|\phi|^2$ has to have a minimum at $|\phi|\equiv \phi_0/\sqrt2$ such that
\begin{align}
0 \leq \sqrt{\frac{2 U(\phi_0/\sqrt2)}{\phi_0^2}} \equiv \omega_0 < m_\phi
\label{eq:minimum}
\end{align}
for an attractive force to exist that leads to large Q-balls~\cite{Coleman:1985ki}. Under these conditions, spherically-symmetric localized solutions to the classical equations of motion of the form $\phi (\vec{x},t) =  f(|\vec{x}|) e^{\ii \omega t}\phi_0/\sqrt2$ exist if 
\begin{align}
    \omega_0 < \omega < m_\phi\,,
    \label{eq:omega_range}
\end{align}
which describe Q-balls~\cite{Coleman:1985ki}.

The attractive-force requirement from Eq.~\eqref{eq:minimum} cannot be satisfied for a renormalizable bounded-from-below single-field potential~\cite{Coleman:1985ki}. Instead, it is necessary to consider multi-field potentials~\cite{Friedberg:1976me} or higher-dimensional operators obtained by integrating out heavier fields~\cite{Lee:1991ax}. Neglecting quantum corrections, the latter procedure generates polynomial potentials in $|\phi|$. Here, we restrict ourselves to polynomials involving three terms,
\begin{align}
U(|\phi|) = m^2_\phi |\phi|^2 -\beta |\phi|^p+\xi |\phi|^q \,,
\label{eq:phi_potential}
\end{align}
as this is the minimal form that satisfies the requirements for large  Q-balls, also studied in Ref.~\cite{Lennon:2021uqu}. 
The above form for $U$ should cover a large part of physically motivated potentials, at least approximately.
While $p$ and $q$ should be \emph{even integers} for potentials obtained within effective field theory (see Sec.~\ref{sec:UV}), most of our mathematical analysis holds for arbitrary integer or even real exponents satisfying $2 < p < q$. We will show in Sec.~\ref{sec:UV} that multi-field scenarios involving additional \emph{light} fields can lead to odd and even \emph{fractional} $p$ and $q$.

For the above potential, we can calculate the parameters relevant for Eq.~\eqref{eq:minimum} as
\begin{align}
\begin{split}
\phi_0 &= \sqrt2 \left( \frac{(p-2)\beta}{(q-2) \xi} \right)^{\frac{1}{q-p}}\,,\label{eq:phi0}\\
\omega_0 &= \sqrt{m_\phi^2 -\frac{q-p}{q-2}\left(\frac{p-2}{q-2}\right)^{\frac{p-2}{q-p}} \left(\frac{\beta^{q-2}}{\xi^{p-2}}\right)^{\frac{1}{q-p}}} \,,
\end{split}
\end{align}
allowing us to replace the (generally dimensionful) couplings $\beta$ and $\xi$ with the physically relevant $\phi_0$ and $\omega_0$. For $2 < p < q$, $\beta$ and $\xi$ both need to be positive.
The $\beta$ coupling provides the attractive force that enables the bound state and the $\xi$ term keeps the potential bounded from below.
The case $p=4$, $q=6$ has been discussed extensively in the literature, especially in Ref.~\cite{Heeck:2020bau}.

Using the ansatz $\phi (\vec{x},t) =  f(|\vec{x}|) e^{\ii \omega t}\phi_0/\sqrt2$ and rescaling $\vec{x}\to \vec{x}\,\sqrt{m_\phi^2 - \omega_0^2}$ in Eq.~\eqref{eq:lagrange-equation} leads to the equation of motion for the dimensionless function $f(\rho)$, 
\begin{align}
f''(\rho) +\frac{2}{\rho} f'(\rho) + \frac{\dd}{\dd f} V(f) = 0\,,
\label{eq:eom}
\end{align}
$\rho$ being the dimensionless radial coordinate, with effective potential
\begin{align}
V(f) = \frac{(p-q) (\kappa^2-1) f^2 - (q-2) f^p + (p-2) f^q}{2(p-q)}
\label{eq:potential}
\end{align}
and boundary conditions $f'(0)=0$ and $f(\rho\to\infty) =0$. 
We will restrict our analysis to solutions of Eq.~\eqref{eq:eom} with monotonically decreasing non-negative $f$, as these describe the Q-ball ground state configurations~\cite{Friedberg:1976me,Volkov:2002aj,Mai:2012cx,Almumin:2021gax}.
Due to the rescaling, the differential equation~\eqref{eq:eom} only depends on $p$, $q$, and the parameter
\begin{align}
    \kappa^2 \equiv \frac{\omega^2 - \omega_0^2}{m_\phi^2-\omega_0^2}\,, 
    \label{eq:kappa}
\end{align}
which is restricted to $0< \kappa < 1$ from Eq.~\eqref{eq:omega_range} and ultimately determines the Q-ball radius $R$~\cite{Heeck:2020bau}.
The macroscopic Q-ball properties of most interest to us, charge $Q$ and  energy $E$, are given by~\cite{Heeck:2020bau}
\begin{align}
    Q &= \frac{4\pi \phi_0^2\, \omega}{(m_\phi^2 - \omega_0^2)^{3/2}} \, \int_0^\infty \dd \rho\, \rho^2 f^2\,,
\label{eq:charge}\\
E &=\omega Q + \frac{4\pi \phi_0^2}{3\sqrt{m_\phi^2 - \omega_0^2}} \, \int_0^\infty \dd \rho\, \rho^2 {f'}^2\,,
\label{eq:energy}
\end{align}
and thus require knowledge of two dimensionless integrals that are functions of $p$, $q$, and $\kappa$ (or the radius).

\begin{figure}
	\includegraphics[width=0.49\textwidth]{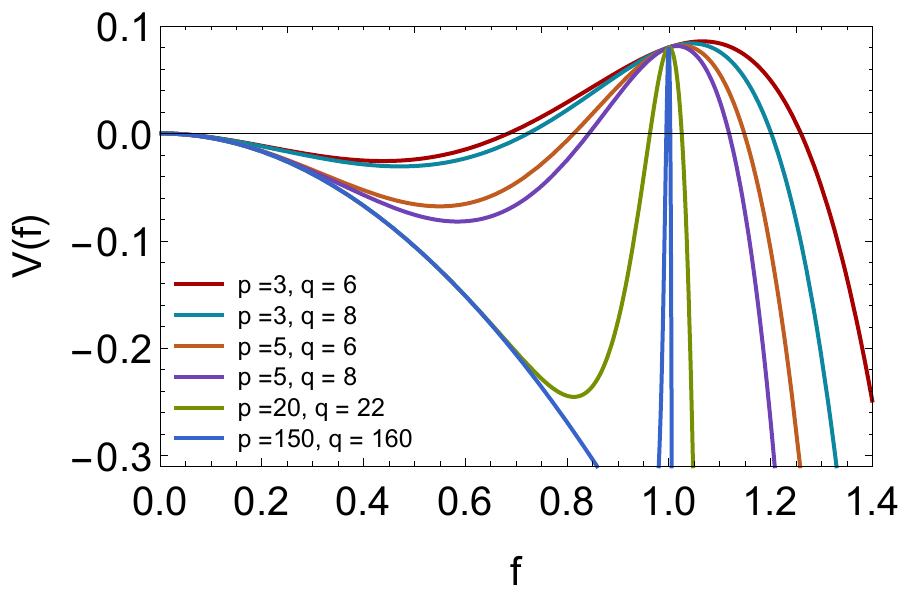}  
	\caption{Plot of the effective potential $V(f)$ from Eq.~\eqref{eq:potential} for $\kappa=0.4$ and various integer $p$ and $q$.
}
	\label{fig:potentialpq}
\end{figure}

Equation~\eqref{eq:eom} can be interpreted as a one-dimensional mechanics problem of a particle with position $f$ moving in the potential $V$, the radial coordinate $\rho$ playing the role of time~\cite{Coleman:1985ki}. In this interpretation, the $f'/\rho$ term corresponds to time-dependent friction.
The potential $V$ is illustrated in Fig.~\ref{fig:potentialpq} for several values of $p$ and $q$. 
For $0< \kappa <1$, the potential has three extrema in the region $f\geq 0$: one local maximum at $f=0$, one local minimum at $f=f_->0$, and a global maximum at $f=f_+ > f_-$. The particle starts at rest at a value $f\in (f_-,f_+)$ and then rolls toward $f=0$, which it reaches after an infinite amount of time, i.e.~for $\rho\to \infty$. For small $\kappa$, $f_+\simeq 1$ and $V(f_+)\simeq \kappa^2/2$, not much larger than $V(0)=0$; the particle therefore needs to start very close to $f_+$ and wait until the friction term is sufficiently suppressed to roll, almost frictionless, to $f=0$. This small $\kappa$ limit is called the thin-wall limit, where $f$ resembles a step function~\cite{Coleman:1985ki} and the Q-ball radius is large. This leading order approximation was investigated in Ref.~\cite{Lennon:2021uqu} as a function of $p$ and $q$.
Following Ref.~\cite{Heeck:2020bau}, we will provide improved approximations for this regime.

\section{Thin-wall limit}
\label{sec:thin-wall}

Neglecting friction in the small-$\kappa$ regime simplifies the equation of motion~\eqref{eq:eom} to
\begin{align}
f''(\rho)  +\left. \frac{\dd}{\dd f} V(f)\right|_{\kappa=0} = 0\,,
\end{align}
which is equivalent to the first-order differential equation
\begin{align}
\frac{1}{2}{f'}^2  +\left. V(f)\right|_{\kappa=0} = 0
\label{eq:mechanical_energy}
\end{align}
upon using energy conservation in the classical-mechanics analogy~\cite{Heeck:2020bau}. The profile $f(\rho)$ is then determined via direct integration as~\cite{Lee:1991ax}
\begin{align}
\left.\int \dd f \, \frac{1}{\sqrt{-2V(f)}} \right|_{\kappa=0}= -\int \dd \rho\,.
\label{eq:transition}
\end{align}
Following Ref.~\cite{Heeck:2020bau}, we denote this solution as the \emph{transition profile}, which is strictly speaking only expected to be valid for small $\kappa$ and around $\rho = R$, but practically provides an excellent approximation for all $\rho$ and even for large $\kappa$, to be specified below.
We define the Q-ball radius $R$ via $f(R)=\tfrac{2}{3} f(0)$, with $f(0)\simeq f_+\simeq 1$ in the thin-wall regime. This proves a more convenient radius definition than that of Ref.~\cite{Heeck:2020bau} as it turns  Eq.~\eqref{eq:transition} into a definite integral that can be calculated numerically with ease to obtain $\rho(f)$~\cite{Lee:1991ax}:
\begin{align}
\rho(f) = R- \left.\int_{2/3}^f \dd \tilde{f} \, \frac{1}{\sqrt{-2V(\tilde{f})}} \right|_{\kappa=0} .
\label{eq:definite_transition}
\end{align}

To estimate the radius $R$ of a Q-ball in the small-$\kappa$ regime we return to the mechanical analogy discussed above. 
The particle starts at $f\simeq f_+\simeq 1$ with potential energy $V(f_+)\simeq \kappa^2/2$ and ends at $f=0$ with potential energy $V(0)=0$. The difference in energies, $\kappa^2/2$, must equal the energy lost through friction~\cite{Mai:2012cx}, i.e.
\begin{align}
\frac{\kappa^2}{2} = -\int_{1}^{0}\dd f \, \frac{2}{\rho}f'(\rho)  \,.
\label{eq:friction_loss}
\end{align}
Since $f'$ is only non-zero around $\rho =R$, we can approximate $1/\rho\simeq 1/R$ in the integrand;
$f'$ can then be replaced by the potential using Eq.~\eqref{eq:mechanical_energy}, giving the relation
\begin{align}
\frac{\kappa^2}{2} \simeq \left. -\frac{2}{R}\int_{1}^{0}\dd f \, \sqrt{-2V(f)}\right|_{\kappa=0} \,.
\label{eq:analog2}
\end{align}
For small $\kappa$, the Q-ball radius is hence of the form
\begin{align}
R \simeq \frac{\eta}{\kappa^2} \equiv \frac{\left. 4\int_{0}^{1}\dd f\, \sqrt{-2 V(f)}\right|_{\kappa=0}  }{\kappa^2}\,.
\label{eq:radius}
\end{align}
As expected from the mechanics analogy, the Q-ball radius becomes larger for decreasing $\kappa$. The prefactor $\eta$ is determined by a simple integral over the potential, which is an $\mathcal{O}(1)$ number with small $p$ and $q$ dependence.  
Using Eq.~\eqref{eq:potential} we can see that the integrand of the Eq.~\eqref{eq:radius} becomes larger with increasing $p$ and $q$. The smallest allowed integers are $p=3$ and $q=4$, which lead to the lower bound $\eta_\text{min}=2/3$. To determine the upper bound of $\eta$, let us take $q$ to infinity first, which leaves us with
\begin{align}
\lim_{q \to \infty} \eta &=4\int_{0}^{1}\dd f \, \sqrt{{f^2-f^p}}\\
&=\frac{\sqrt{\pi }\, \Gamma \left(\frac{p}{p-2}\right)}{\Gamma \left(\frac{3}{2}+\frac{2}{p-2}\right)}
\label{eq:qInfQEta}\\
&= 2-\frac{2.45}{p} + \mathcal{O}\left(p^{-2}\right) ,
\label{eq:qInfQEtaApprox}
\end{align}
 with the Gamma function $\Gamma (x)$. 
From Eq.~\eqref{eq:qInfQEtaApprox} it follows that the upper bound is $\eta_\text{max}=2$. We conclude that for integer exponents $\frac{2}{3}\leq\eta<2$.

Since the potential~\eqref{eq:potential} is symmetric under $p\leftrightarrow q$, subsequent equations should also be symmetric; this leads us to a better approximation for $\eta$:
\begin{align}
\eta &\simeq \frac{\sqrt{\pi }\, \Gamma \left(\frac{p}{p-2}\right)}{\Gamma \left(\frac{3}{2}+\frac{2}{p-2}\right)}+\frac{\sqrt{\pi }\, \Gamma \left(\frac{q}{q-2}\right)}{\Gamma \left(\frac{3}{2}+\frac{2}{q-2}\right)} -2\,,
\end{align}
valid for large $p$ and $q$. This deviates from the numerical integral of Eq.~\eqref{eq:radius} by less than 8\% for integer $p>3$ and $q>6$ and is therefore a useful approximation for most exponents.

In the thin-wall limit, $\kappa\ll 1$, the Q-ball radius is hence obtained from Eq.~\eqref{eq:radius} and the profile $f(\rho)$ -- or rather $\rho(f)$ -- from Eq.~\eqref{eq:definite_transition}, which can then be used to obtain Q-ball charge and energy from Eqs.~\eqref{eq:charge} and~\eqref{eq:energy} using the integrals
\begin{align}
 \int\limits_0^\infty \dd \rho\, \rho^2 f^2 &\simeq \int\limits_0^{\rho(1-\epsilon)}\dd\bar{\rho} \,\bar{\rho}^2 + \int\limits_0^{1-\epsilon}\dd f\, \frac{\rho(f)^2 f^2}{\sqrt{\left. -2 V(f)\right|_{\kappa=0}}}\,,
\\
\int\limits_0^\infty \dd \rho\, \rho^2 {f'}^2 &\simeq \int\limits_0^{1}\dd f\, \rho(f)^2 \sqrt{\left.-2 V(f)\right|_{\kappa=0}}\,,
\end{align}
where the $f^2$ integral is split to avoid the singularity in the second term for $\epsilon\to 0$. Any $\epsilon\ll 1$ gives a good approximation here.
This procedure is trivial to perform numerically for any $p$ and $q$ and is far simpler than solving the original differential equation, especially since the latter becomes numerically difficult for minuscule $\kappa$.
For some cases of $p$ and $q$, all integrals can even be performed analytically, leading to particularly simple descriptions of thin-wall Q-balls, as shown below.

To compare our analytic approximations with the exact solutions, we solve the differential equation~\eqref{eq:eom}  numerically via the shooting method~\cite{Coleman:1985ki}, which is straightforward at least for small $p$ and $q$ and $\kappa$ not too close to $0$ or $1$. Since the differential equation including boundary conditions is identical to the bounce equation of vacuum decay in three dimensions~\cite{Coleman:1977py,Callan:1977pt,Coleman:1977th}, we can borrow codes dedicated to that problem to find Q-ball profiles. In addition to our own implementation of the shooting method, we also use \texttt{AnyBubble}~\cite{Masoumi:2016wot} in our analysis.

\subsection{Equidistant exponents: \texorpdfstring{$p=2+n$, $q=2+2n$}{p=2+n, q=2+2n}}
\label{sec:equidistant}

\begin{figure}[tb]
	\includegraphics[width=0.48\textwidth]{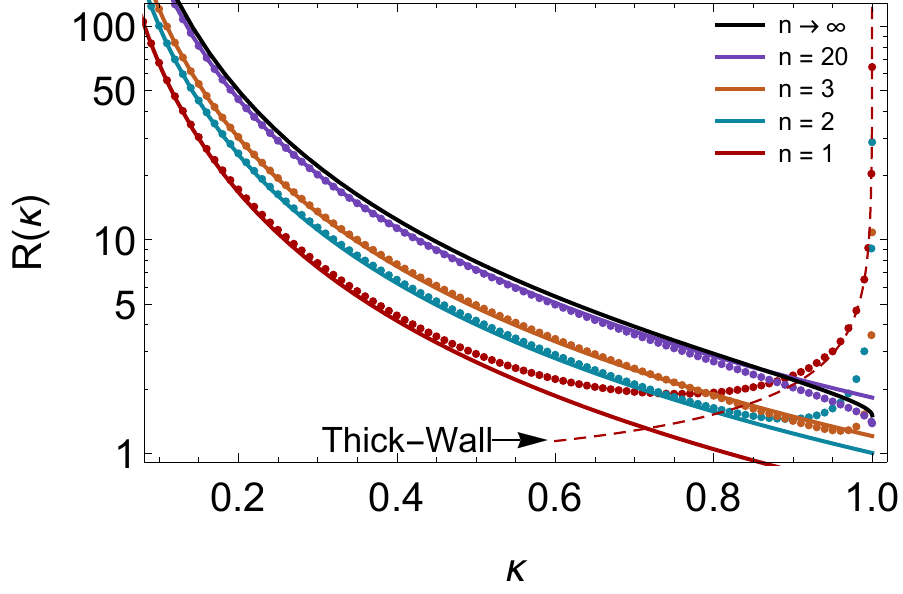}  
	\caption{$R(\kappa)$ dependence for various equidistant exponents, $p=2+n$, $q=2+2n$. Dots correspond to numerical values and solid lines to the $R\propto 1/\kappa^2$ thin-wall prediction of Eq.~\eqref{eq:radius_n}. The dashed line represents the $n=1$ thick-wall prediction of Sec.~\ref{sec:thick} and the black line the $n\to\infty$ limit from Sec.~\ref{sec:largen}.
}
	\label{Plot:RvsKMod}
\end{figure}

Analytic approximations of the thin-wall Q-ball equation are easiest to obtain when the exponents in the potential are equidistant, i.e.~$p-2=q-p \equiv n$, where $n$ is positive and typically an even integer. 
Special cases include $n=2$, discussed in Ref.~\cite{Heeck:2020bau}, and $n=1$, discussed in Ref.~\cite{Kusenko:1997ad}.
In this case, the potential $V$ reaches its global maximum at
\begin{align}
f_+ &= \left(\frac{2+n+\sqrt{n^2 + 4 \kappa^2 + 4 n \kappa^2}}{2+2n}\right)^{\frac{1}{n}}\\
&= 1 + \frac{\kappa^2}{n^2} - \frac{(1+3n)\kappa^4}{2n^4} + \mathcal{O}(\kappa^6)\,.
\label{eq:fplus}
\end{align}
The magnitude of the potential at this point is
\begin{align}
V(f_+) = \frac{\kappa^2}{2}\left[ 1 + \frac{\kappa^2}{n^2} - \frac{\kappa^4}{n^3} + \mathcal{O}(\kappa^6) \right] .
\label{eq:Vfplus}
\end{align}
The radius integral~\eqref{eq:radius} can be performed analytically to give the radius at small $\kappa$:
\begin{align}
    R \simeq \frac{2 n}{(2 + n)\kappa^2}\,,
    && \eta \simeq \frac{2 n}{2 + n}\,.
    \label{eq:radius_n}
\end{align}
Restricting ourselves to integer $n$, the coefficient $\eta$ ranges from $2/3$ ($n=1$) to $\eta=2$ ($n\to \infty$), increasing monotonically. This happens to coincide with the $\eta$ range for arbitrary $p$ and $q$, as shown above.
In Fig.~\ref{Plot:RvsKMod}, we compare the prediction from Eq.~\eqref{eq:radius_n} with numerical results\footnote{Numerical data are supplied as ancillary files on the arXiv~\cite{Heeck:2022iky}.\label{ancillary_files}} for several $n$ and find excellent agreement even for $\kappa$ as large as $0.8$. The only exception is the $n=1$ case, which is special in many ways and will be discussed in more detail below.

The analytical transition function of Eq.~\eqref{eq:definite_transition} takes the simple form
\begin{align}
f(\rho)=\left[1+\bigg(\bigg[\frac{3}{2}\bigg]^n-1\bigg)\, e^{n(\rho-R)}\right]^{-\frac{1}{n}}.
\label{eq:profile2Mod}
\end{align}
(For the radius definition of Ref.~\cite{Heeck:2020bau}, $f''(R)=0$, we would instead have $f(\rho) = \left(1+n \exp[n(\rho-R)]\right)^{-1/n}$.)

\begin{figure}[tb]
	\includegraphics[width=0.48\textwidth]{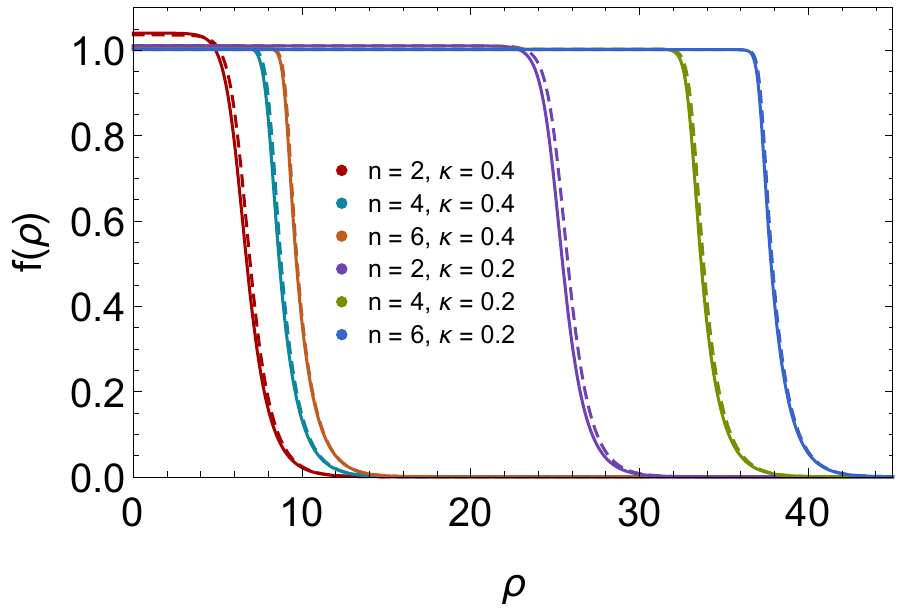}  
	\caption{Comparison between profiles generated numerically (dashed) and analytically (solid) via Eqs.~\eqref{eq:transitionF} and~\eqref{eq:radius_n}.
}
	\label{Plot:Profiles1Mod}
\end{figure}

Rather than using this transition profile $f(\rho)$ directly, we modify it slightly to take into account that the particle does not start at $f=1$ but rather $f_+\simeq 1+\kappa^2/n^2$ and define
\begin{align}
F(\rho)\equiv  \left(1+\frac{\kappa^2}{n^2}\right) \left[1+\bigg(\bigg[\frac{3}{2}\bigg]^n-1\bigg)\, e^{n(\rho-R)}\right]^{-\frac{1}{n}}\,.
\label{eq:transitionF}
\end{align}
This ansatz $F(\rho)$ is equally valid as $f(\rho)$ but leads to a slightly better agreement with numerical results for larger $\kappa$.
In Fig.~\ref{Plot:Profiles1Mod} we can see how well this approximation describes the exact profiles. The transition profiles become better for smaller $\kappa$, as expected, as well as for larger $n$. The latter can be understood by noting that our thin-wall approximations $f_+ \simeq 1$ and $V(f_+)\simeq \kappa^2/2$ become increasingly better for larger $n$, as can be seen in Eqs.~\eqref{eq:fplus} and~\eqref{eq:Vfplus}.

With radius and transition profiles at our disposal it is straightforward to calculate Q-ball charge and energy, as determined by the two integrals $\int  F^2\rho^2\dd \rho$ and $\int  F'^2\rho^2\dd \rho$. Expanding in small $\kappa$ or large radius, we find
\begin{widetext}
\begin{align}
\begin{split}
&\int_{}^{} [F(\rho)]^2\rho^2\dd\rho \simeq \frac{R^3}{3}\left[1+\frac{1}{n (n+2) R} \left(-3 (n+2) \log \left(\left(\frac{3}{2}\right)^n-1\right)-3 (n+2) \left(\psi ^{(0)}\left(\frac{2}{n}\right)+\gamma \right)+4\right)\right.\\
&+\frac{1}{2 n^2 R^2}\left\{\pi ^2 + 6 \gamma  \left(\gamma -\frac{4}{n+2}\right)+12 \left(\log \left(\left(\frac{3}{2}\right)^n-1\right)+\gamma -\frac{2}{n+2}\right) \psi ^{(0)}\left(\frac{2}{n}\right)
\right.\\
&\left.\left.+\frac{6 \log \left(\left(\frac{3}{2}\right)^n-1\right) \left(2 \gamma  (n+2)+(n+2) \log \left(\left(\frac{3}{2}\right)^n-1\right)-4\right)}{n+2}+\frac{8}{(n+2)^2}+6 \psi ^{(0)}\left(\frac{2}{n}\right)^2+6 \psi ^{(1)}\left(\frac{2}{n}\right)\right\}\right] ,
\end{split}
\label{eq:intF}\\
\begin{split}
&\int_{}^{} [F'(\rho)]^2\rho^2\dd\rho 
\simeq\frac{nR^2}{2n+4}\left[1+\frac{4-2 (n+2) \left(\log \left(\left(\frac{3}{2}\right)^n-1\right)+\psi ^{(0)}\left(\frac{2}{n}\right)+\gamma -1\right)}{n (n+2) R}\right.\\
&+\frac{1}{6 n^2 (n+2)^2 R^2}\left\{48 (n+2) \left(\log \left(\left[\frac{3}{2}\right]^n-1\right)+\psi ^{(0)}\left[\frac{2}{n}\right]+\gamma -1\right)\right.\\
&+24+(n+2)^2\left( 6 \psi ^{(1)}\left[\frac{2}{n}\right]+\pi ^2+6 (\gamma -2) \gamma+6 \log \left(\left[\frac{3}{2}\right]^n-1\right) \right.\left(\log \left(\left[\frac{3}{2}\right]^n-1\right)+2 \gamma -2\right)\\
&\left.\left.\left. +6 \psi ^{(0)}\left[\frac{2}{n}\right] \left(2 \left(\log \left(\left[\frac{3}{2}\right]^n-1\right)+\gamma -1\right)+\psi
   ^{(0)}\left[\frac{2}{n}\right]\right)\right)\right\}\right] ,
\end{split}
\label{eq:intFP}
\end{align}
\end{widetext}
where $\gamma\simeq 0.577$ is the Euler--Mascheroni constant and $\psi^{(1)}(x)$ is the first derivative of the Digamma function $\psi^{(0)} (x)\equiv \Gamma'(x)/\Gamma(x)$.

\begin{figure}[tb]
	\includegraphics[width=0.48\textwidth]{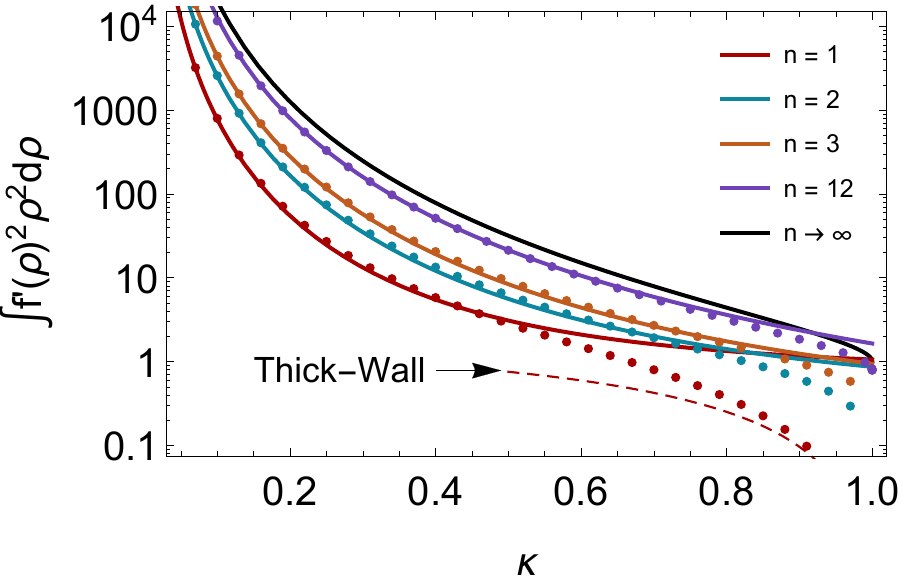}\\
	\includegraphics[width=0.48\textwidth]{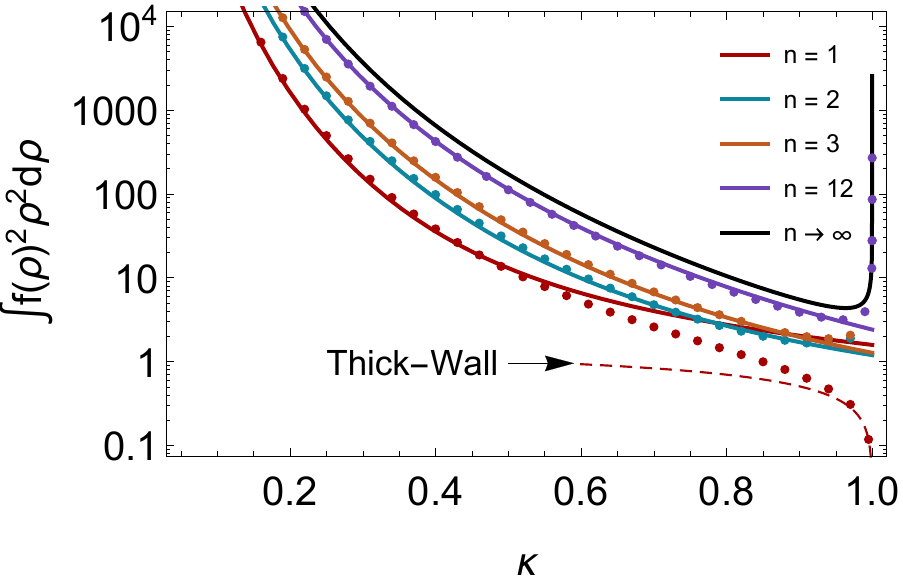}  
	\caption{$\int [F'(\rho)]^2\rho^2\dd\rho $ (top) and $\int [F(\rho)]^2\rho^2\dd\rho $ (bottom) as functions of $\kappa$ for various $n$. Dots correspond to the numerical values, solid lines show our thin-wall approximations from Eqs.~\eqref{eq:intF} and~\eqref{eq:intFP}.
 The dashed line represents the $n=1$ thick-wall prediction of Sec.~\ref{sec:thick} and the black line the $n\to\infty$ limit from Sec.~\ref{sec:largen}.
}
	\label{Plot:Integrals_n}
\end{figure}

In figure~\ref{Plot:Integrals_n} we compare these integrals to the numerical solutions for various $n$. 
Clearly our analytical approximations are excellent even outside the thin-wall limit. For $n>1$, they are good up to $\kappa\simeq 0.8$ and become better for increasing $n$. The case $n=1$ is once again special and will be discussed in more detail below in section~\ref{sec:thick}.

The two integrals allow us to determine the charge $Q$ and energy $E$ of Q-balls. To lowest non-trivial order we have
\begin{align}
    E \simeq \omega_0 Q + \frac{n}{2+n}\left( \frac{9\pi \phi_0^2}{2\omega_0^2}\right)^{1/3}\sqrt{m_\phi^2-\omega_0^2}\, Q^{2/3}
\end{align}
in the thin-wall or large-$Q$ limit $\omega\simeq \omega_0$, assuming $\omega_0 \neq 0$.\footnote{For $\omega_0=0$, we have $E\simeq \tfrac{5}{2} [n/(2+n)]^{3/5}(\pi/3)^{1/5}\phi_0^{2/5}m_\phi^{3/5} Q^{4/5}$.} We see that the $n$-dependence of the Q-ball energy for a fixed charge $Q$ is very mild, merely an $\mathcal{O}(1)$ factor in front of the surface energy.
Of particular interest is the ratio $E/(m_\phi Q)$, which has to be smaller than unity to ensure Q-ball stability against decay into $Q$ free particles~\cite{Tsumagari:2008bv}:
\begin{align}
\begin{split}
\frac{E}{m_\phi Q}&=\sqrt{\kappa^2+\Big[\frac{\omega_0}{m_\phi}\Big]^2(1-\kappa^2)}\\
&\quad +\frac{1-\big[\frac{\omega_0}{m_\phi}\big]^2}{3\sqrt{\kappa^2+\big[\frac{\omega_0}{m_\phi}\big]^2(1-\kappa^2)}}\,\frac{\int_{}^{} [f'(\rho)]^2\rho^2\dd\rho }{\int_{}^{} [f(\rho)]^2\rho^2\dd\rho }\,.
\end{split}
\label{eq:EmQ}
\end{align}
The  stability criterion $E/(m_\phi Q)<1$ depends on $\kappa$, $\omega_0/m_\phi$, and the ratio of the two integrals. In the small $\kappa$ expansion, the ratio of integrals takes the following form
\begin{align}
\begin{split}
&\frac{\int_{}^{} [F'(\rho)]^2\rho^2\dd\rho \,}{\int_{}^{}[F(\rho)]^2\rho^2\dd\rho \,} \simeq\frac{3n}{2(2+n)R}\\
&\times \left[1+\frac{1}{R}\frac{2+\gamma+\ln{\left[\left(\frac{3}{2}\right)^n-1\right]}+\psi^{(0)} (\frac{2}{n})}{n}+\frac{1}{3n^2R^2}\right.\\
&\times\left\{3\ln{\left[\left(\frac{3}{2}\right)^n-1\right]}\left[4+2\gamma+\ln{\left[\left(\frac{3}{2}\right)^n-1\right]}\right]\right.\\
&+6\psi^{(0)} \left(\frac{2}{n}\right)\left[2+\gamma+\ln{\left[\left(\frac{3}{2}\right)^n-1\right]}+\frac{\psi^{(0)} \left(\frac{2}{n}\right)}{2}\right]\\
&\left.\left.+3\gamma(4+\gamma)-\pi^2-6\psi^{(1)}\left(\frac{2}{n}\right)  \right\}\right] .
\end{split}
\label{eq.Ratio}
\end{align}
For small $\kappa$ or large $R$, the ratio of integrals goes to zero as $3\kappa^2/4$ and $E\to \omega_0 Q< m_\phi Q$ for $\omega_0>0$. Stability against decay into $Q$ free particles is hence guaranteed in the thin-wall limit, as shown long ago by Coleman~\cite{Coleman:1985ki}. This holds for all $n$.

\begin{figure}[tb]

	\includegraphics[width=0.48\textwidth]{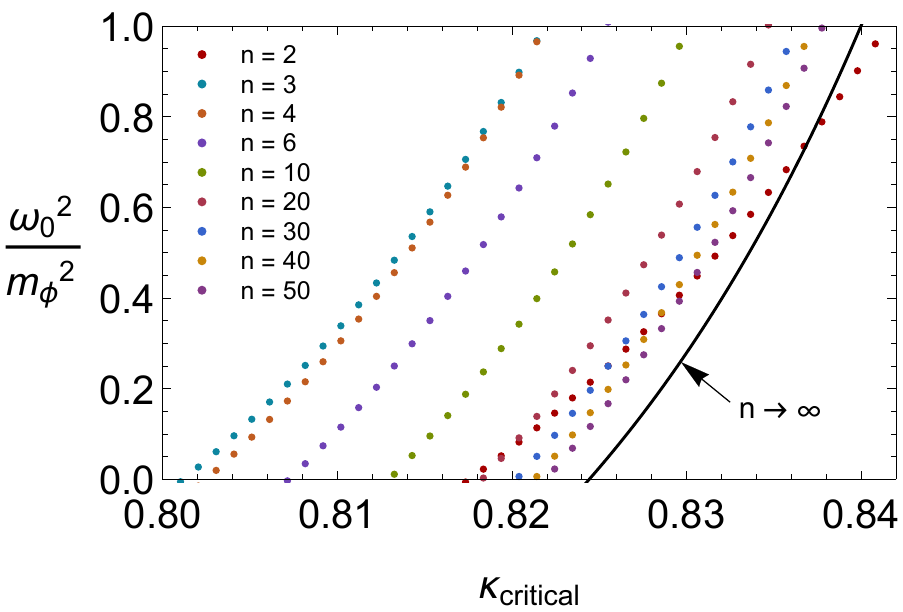}  
	\caption{Values of $\kappa=\kappa_\text{critical}$ that leads to $E=m_\phi Q$ for several $n$.
	Dotted curves are generated numerically, the solid black line represents our theoretical estimation of the $n\to\infty$ case from Sec.~\ref{sec:largen}.
}
	\label{Plot:PlotEmQEquidistant}
\end{figure}

For larger $\kappa$, on the other hand, it is not clear that $E$ remains below $m_\phi Q$; indeed, our analytical thin-wall results imply $E> m_\phi Q$ for $\kappa \gtrsim 0.8$ for all integer $n$. Unfortunately, this $\kappa$ region is just at the edge of viability for our thin-wall results and hence not fully trustworthy, at least for small $n$.
Instead, we have checked this region numerically; for $n\geq 2$, Q-balls indeed become unstable for $\kappa \geq \kappa_\text{critical} \sim 0.8$, illustrated in Fig.~\ref{Plot:PlotEmQEquidistant}. For $n\geq 3$, a regular pattern emerges where $\kappa_\text{critical}$ increases with $n$. This eventually converges toward the black $n\to\infty$ line in Fig.~\ref{Plot:PlotEmQEquidistant}, which is derived in Sec.~\ref{sec:largen} and does not rely on the thin-wall approximation.
Ultimately, $\kappa_\text{critical}$ always lies between $0.8$ and $0.85$ for $n\geq 2$, showing a rather mild dependence on $n$ and $\omega_0/m_\phi$. 
Since $Q\propto \int \dd\rho \,\rho^2f^2$ is a monotonic function of $\kappa$ for $\kappa< \kappa_\text{critical}$ (see Fig.~\ref{Plot:Integrals_n}), the stability criterion is equivalent to a minimal charge $Q$ a stable Q-ball needs to have.

\begin{figure}[tb]
	\includegraphics[width=0.48\textwidth]{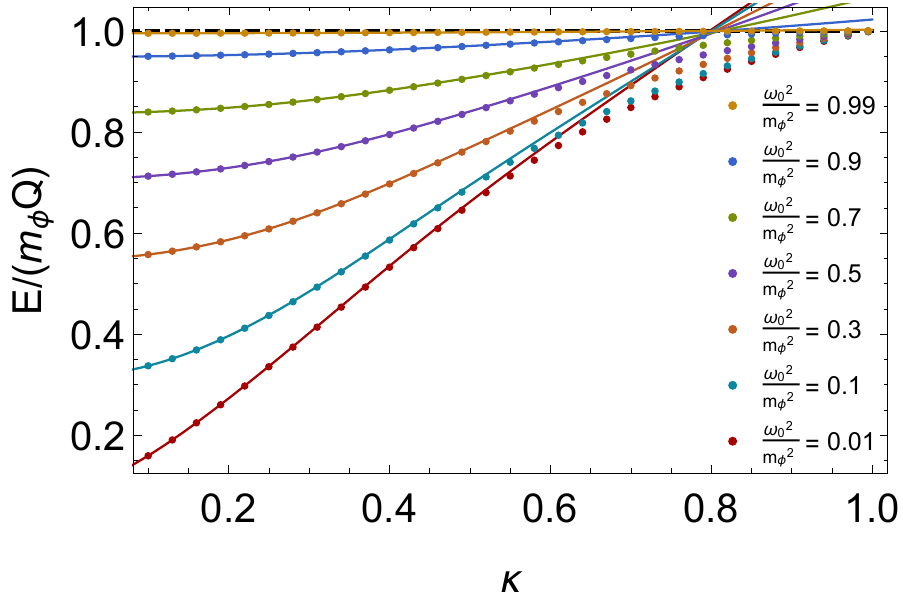}  
	\caption{$E/(m_\phi Q)$ dependence on $\kappa$ for $n=1$ for various $\omega _0^2/m_\phi^2$.  The dots represent numerical data, the lines our thin-wall results.
}
	\label{Plot:EmQp3q4}
\end{figure}

For integer $n$, we are left with the special case $n=1$ (or $p=3$, $q=4$), which does not have a $\kappa_\text{critical}$, i.e.~leads to Q-balls with $E< m_\phi Q$ for all $\kappa\in (0,1)$, see Fig.~\ref{Plot:EmQp3q4}. The stability of these Q-balls around $\kappa\sim 1$ was proven already in Refs.~\cite{Kusenko:1997ad,PaccettiCorreia:2001wtt,Sakai:2007ft,Lennon:2021uqu} (see Sec.~\ref{sec:thick} for details), here we show numerically that the Q-balls are also stable in the intermediate $\kappa$ regime between the thin- and thick-wall limits. Analytical approximations are difficult to obtain in this intermediate regime.

\subsection{General exponents}

Equidistant exponents in the potential lead to simple analytical expressions for thin-wall Q-ball properties, but clearly only cover part of the possible parameter space. Let us briefly discuss general $p$ and $q$ exponents.
Notice that even though our original Lagrangian requires $p<q$, the rescaled differential equation and effective potential $V(f)$  are symmetric under $p\leftrightarrow q$ and thus equally valid for $q>p$; even the limit $q=p$ is well defined.

\begin{figure}[tb]
	\includegraphics[width=0.48\textwidth]{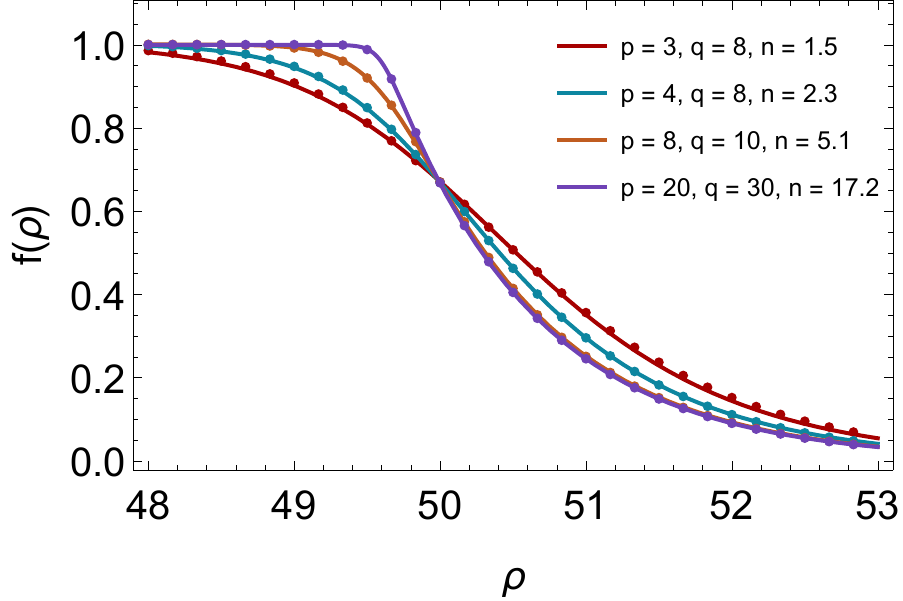}  
	\caption{Profile behavior in the vicinity of the surface for $R=50$. Solid lines correspond to the theoretical prediction of Eq.~\eqref{eq:transitionF} with effective $n(p,q)$ from Eq.~\eqref{eq:NPQ}, while dots come from numerical computation.
}
	\label{Plot:ProfilesRad50}
\end{figure}

Eq.~\eqref{eq:definite_transition} cannot be solved analytically for arbitrary $p$ and $q$, but we can try to find an \emph{effective} equidistance parameter $n(p,q)$ which generates the profile most similar to the one generated by $p$ and $q$. To find this $n$, we note that both radius and profile shape are determined by integrals of the function $\sqrt{-2 V(f)}|_{\kappa=0}$, see Eqs.~\eqref{eq:radius} and~\eqref{eq:definite_transition}. This function $\sqrt{-2 V(f)}|_{\kappa=0}$ is fairly simple: it vanishes at $f=0$ and $f=1$ and has a $(p,q)$-dependent maximum at an $f\in (1/2,1)$. For any $p$ and $q$ we can try to describe this function approximately using the equidistant expression $\sqrt{-2 V(f)}|_{\kappa=0} = f(1-f^n)$. A numerical fit would lead to the optimal $n$, but to obtain an \emph{analytic} approximation we simply match the potentials at the radius, i.e.~at $f=2/3$:
\begin{align}
    V\left(f=\frac{2}{3}\right)_{p=2+n,q=2+2n}=V\left(f=\frac{2}{3}\right)_{p,q}
\end{align}
which provides the effective  $n(p,q)$
\begin{align}
n(p,q)=\frac{\log \left(1-\frac{3}{2} \sqrt{\frac{\left(\frac{2}{3}\right)^q(p-2)
   }{q-p}+\frac{\left(\frac{2}{3}\right)^p
   (q-2)}{p-q}+\frac{4}{9}}\right)}{\log \left(\frac{2}{3}\right)},
\label{eq:NPQ}
\end{align}
manifestly symmetric under $p\leftrightarrow q$.
This ansatz for $n(p,q)$ can now be used with Eq.~\eqref{eq:transitionF} to predict the profile $f(\rho)$ for arbitrary $p$ and $q$.
We stress that the so-obtained $f(\rho)$ will always be approximate, unlike the equidistant cases that correspond to actual asymptotic solutions to the differential equation. Nevertheless, the profile obtained using this effective $n(p,q)$ is a good approximation of the actual potential $V(f,p,q)$, especially for $p\simeq q$.
Profiles generated using this $n(p,q)$ prediction and Eq.~\eqref{eq:transitionF} can be seen in Fig.~\ref{Plot:ProfilesRad50}.
The one-parameter set of profiles of Eq.~\eqref{eq:transitionF} is apparently (and surprisingly) sufficient to capture  all possible profile shapes for general exponents $p$ and $q$!

\begin{figure}[tb]
	\includegraphics[width=0.48\textwidth]{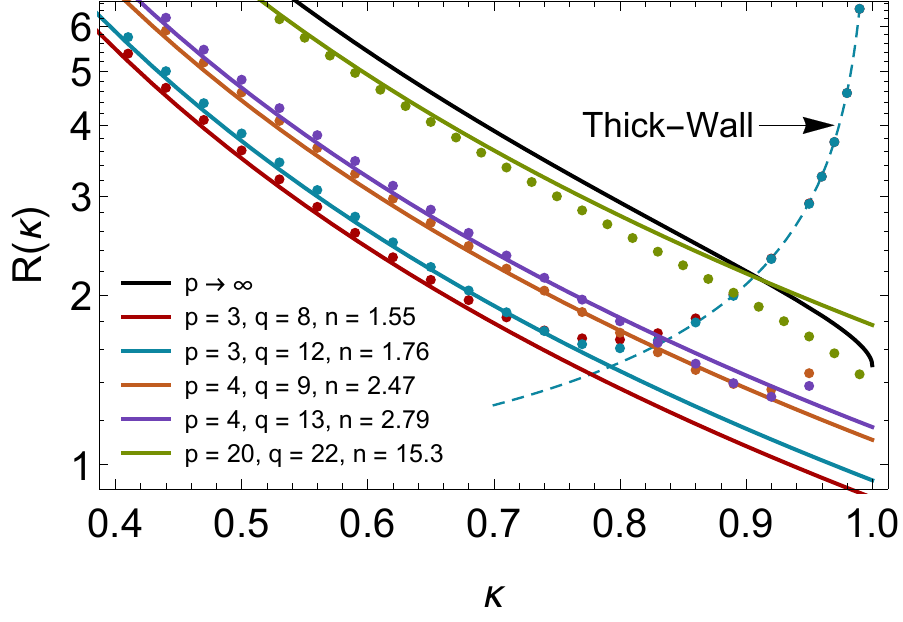}  
	\caption{$R(\kappa)$ dependence for various $p$ and $q$ using $n(p,q)$ prediction. Dots correspond to the numerical values, solid lines show $R=2n/(2+n)/\kappa^2$ with $n(p,q)$ from Eq.~\eqref{eq:NPQ}. The dashed line represents the thick-wall prediction for $p=3$ derived in Sec.~\ref{sec:thick} and the black line the $n\to\infty$ limit from Sec.~\ref{sec:largen}.}
	\label{Plot:RvsKNPQ}
\end{figure}

This $n(p,q)$ prediction naturally allows us to apply \emph{every} analytical formula that we have already derived for the equidistant case in Sec.~\ref{sec:equidistant} to the general case of arbitrary $(p,q)$. For example, with the help of equations ~\eqref{eq:radius_n} and~\eqref{eq:NPQ} we can predict radii of Q-balls for arbitrary $p$ and $q$. In Fig.~\ref{Plot:RvsKNPQ} we can see that we predict radii with high accuracy all the way up to $\kappa\simeq 0.8$, at least for $p>3$. The region beyond $\kappa\approx 0.86$ is unstable anyway in all cases except $p=3$, to be discussed in detail in Sec.~\ref{sec:thick}. It is worth noting that $R(\kappa)$ calculated in this way and using the previous prediction (Eq.~\eqref{eq:radius}) agree to better than 7\% for all integer $p$ and $q$, the largest deviation being $6.3\%$ for the $p=5$, $q\to\infty$ case.

\begin{figure}[tb]
	\includegraphics[width=0.48\textwidth]{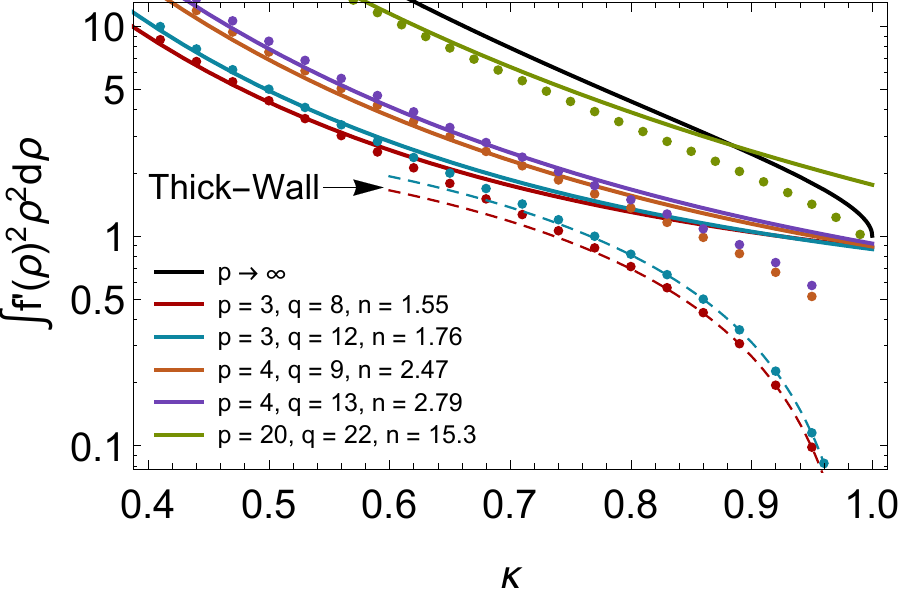}  \\
	\includegraphics[width=0.48\textwidth]{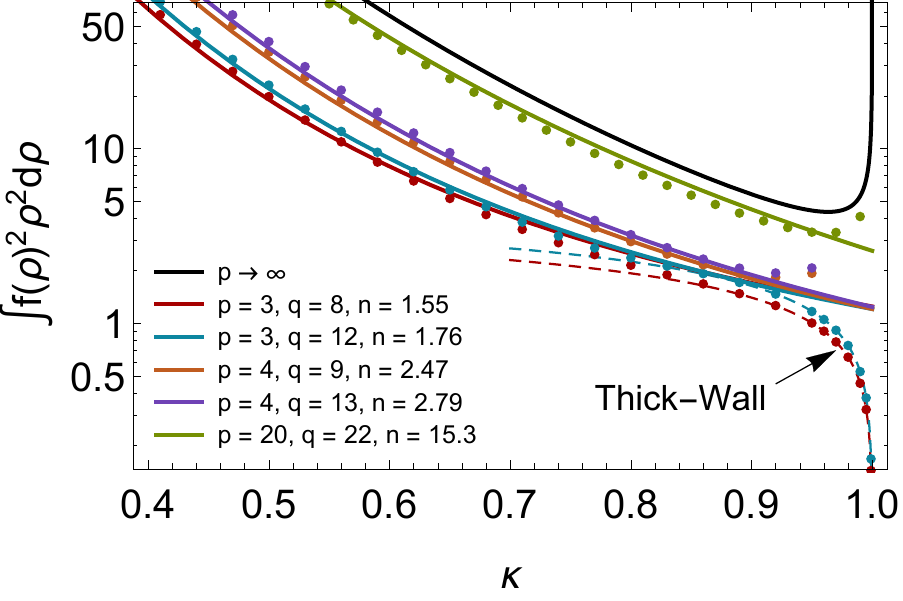}  
	\caption{$\int_{}^{} [F'(\rho)]^2\rho^2\dd\rho \,$ (top) and $\int_{}^{} [F(\rho)]^2\rho^2\dd\rho \,$ (bottom) dependence on $\kappa$ for various $p$ and $q$ using $n(p,q)$ prediction. Dots correspond to the numerical values, when solid lines denote the predictions from Eqs.~\eqref{eq:intFP} and~\eqref{eq:intF}. The dashed lines represent the thick-wall predictions for $p=3$ derived in Sec.~\ref{sec:thick}  and the black lines the $n\to\infty$ limit from Sec.~\ref{sec:largen}.
}
	\label{Plot:IntvsKNPQ}
 
\end{figure}

With radius and profile for arbitrary $p$ and $q$ at our disposal, we can also calculate the two integrals relevant for Q-ball energy and charge. The integrals are simply Eqs.~\eqref{eq:intF} and~\eqref{eq:intFP}, with the radius replaced by Eq.~\eqref{eq:radius_n} and $n$ by the effective $n(p,q)$ from Eq.~\eqref{eq:NPQ}. The comparison to numerical results is shown in Fig.~\ref{Plot:IntvsKNPQ} and is very good for small $\kappa$ and $p>3$. We can see that $\int_{}^{} [F(\rho)]^2\rho^2\dd\rho \,$ (bottom) works extremely well for $\kappa\lesssim0.86$. 
$\int_{}^{} [F'(\rho)]^2\rho^2\dd\rho \,$ (top) properly fits numerical results only for $\kappa\lesssim0.75$. The $p=3$ case is special as Q-balls remain stable for all $\kappa$ and also shows the largest deviation with our prediction. This case is discussed in more details in Sec.~\ref{sec:thick}.

\begin{figure}[tb]
	\includegraphics[width=0.48\textwidth]{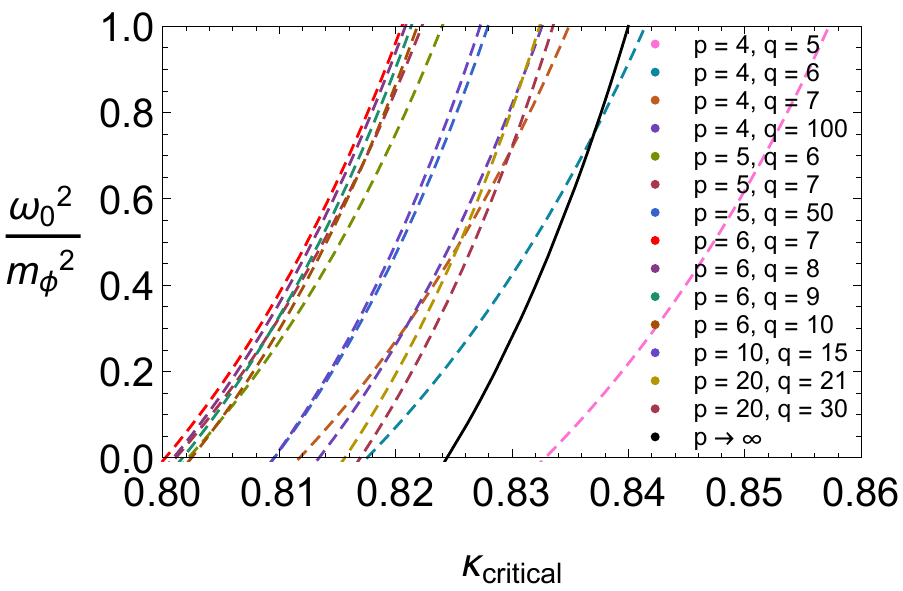}
	\caption{ $\kappa_\text{critical}$ (where $E=m_\phi Q$) for several $p$ and $q$.
}
	\label{Plot:EmQAll}
\end{figure}

Just like in the case of equidistant exponents, our thin-wall results predict $E>m_\phi Q$ for $\kappa\gtrsim 0.8$. For large $p$ and $q$, this region is still covered by our thin-wall approximation and hence qualitatively correct. For small $p$ and $q$, we have to rely on numerical data to investigate Q-ball stability in this region. 
As shown already in Refs.~\cite{PaccettiCorreia:2001wtt,Sakai:2007ft,Lennon:2021uqu}, only the cases with $p=3$ are stable near $\kappa=1$, and are actually stable for all $\kappa$, as argued below in Sec.~\ref{sec:thick}. All integer cases with $p>3$, on the other hand, become unstable beyond some $\kappa_\text{critical}\sim 0.8$. 
We provide many examples for $\kappa_\text{critical}$ in Fig.~\ref{Plot:EmQAll}.
Our numerical calculations show that the case of $p=4$ \& $q=5$ has the largest $\kappa_\text{critical}$ compared to other integer exponents. As we increase both $p$ and $q$ we see that $\kappa_\text{critical}$ decreases. And as we keep increasing $p$ and $q$, at some point $\kappa_\text{critical}$ starts increasing again. We find that out of all integer exponents $p=6$ \& $q=7$ case has the smallest $\kappa_\text{critical}$. Also, in this case and in all the following cases with larger $p$, increasing $q$ increases $\kappa_\text{critical}$ as well. If we take a look at larger exponents we can notice that curves follow more consistent shape and look similar to our $p\to\infty$ prediction. We derived this prediction using Eq.~\eqref{eq:large_p_integrals} and Eq.~\eqref{eq:EmQ}.

\section{Thick-wall limit and the \texorpdfstring{$p=3$}{p=3} case}
\label{sec:thick}

As shown above, solutions to the differential equation~\eqref{eq:eom} can be well approximated in the small $\kappa$ regime using transition functions. For integer $p>3$, these approximations are sufficiently accurate over the entire $\kappa$ region that leads to stable Q-balls.
Only the cases with $p=3$ motivate us to consider larger $\kappa$, as they still allow for stable Q-balls~\cite{Kusenko:1997ad,PaccettiCorreia:2001wtt,Sakai:2007ft,Lennon:2021uqu}.
Near $\kappa\sim 1$ we can find the profiles using the thick-wall approximation~\cite{Kusenko:1997ad,Linde:1981zj}, based on the fact that $f(0)$ becomes smaller and smaller as $\kappa\to 1$, which is clear from the shape of the potential $V$. For small $f$, one can then neglect the $f^q$ term in the potential, seeing as it is the most suppressed term in the small-$f$ limit. In the classical-mechanics analogy, the particle is not starting near the maximum as in the thin-wall limit, so the $f^q$ term that generates this maximum can be neglected.
Notice that we still have a $q$-dependence in our potential despite neglecting the $f^q$ term due to our definitions of e.g.~$\omega_0$. $q$ of course drops out of physical quantities in the thick-wall limit. Setting $p=3$ and omitting $f^q$ allows for a useful rescaling of the differential equation~\cite{PaccettiCorreia:2001wtt}: we write 
\begin{align}
f(\rho) = \frac{2(q-3)}{3(q-2)}(1-\kappa^2) \,\, g\left( \sqrt{1-\kappa^2} \rho\right)
\end{align}
with a function $g(x)$ that is determined by the parameterless differential equation
\begin{align}
g''(x) +\frac{2}{x} g'(x) -g(x) + g(x)^2 = 0\,,
\end{align}
easily solved numerically\footref{ancillary_files} and well-approximated by the function
\begin{align}
g(x)\simeq  (4.20 - 0.10 x - 0.85 x^2 +  0.30 x^3)e^{-0.31 x^2}\,.
\end{align}
The Q-ball radius in the thick-wall limit then diverges as
\begin{align}
R \simeq \frac{0.91}{\sqrt{1-\kappa^2}}
\quad \Rightarrow\quad
R_\text{Q-ball} \simeq \frac{0.91}{\sqrt{m_\phi^2-\omega^2}}
\end{align}
and the integrals take the simple form
\begin{align}
\int_{}^{} [f'(\rho)]^2\rho^2\dd\rho
&= (1-\kappa^2)\int_{}^{} [f(\rho)]^2\rho^2\dd\rho\\
&= \frac{4(q-3)^2}{9(q-2)^2}(1-\kappa^2)^{\frac{3}{2}}\underbrace{\int \dd x \,x^2 g^2}_{\simeq 10.42} .
\end{align}
These thick-wall predictions are shown in Figs.~\ref{Plot:RvsKMod}, \ref{Plot:Integrals_n},  \ref{Plot:RvsKNPQ}, and~\ref{Plot:IntvsKNPQ}  and match numerical data very well for $\kappa$ close to $1$, especially for $q\gg 3$.
For the stability ratio we then find
\begin{align}
\frac{E}{m_\phi Q} = 1 - \frac{m_\phi^2-\omega_0^2}{3m_\phi^2}(1-\kappa^2) + \mathcal{O}\left[(1-\kappa^2)^2\right],
\end{align}
rendering these $p=3$ thick Q-balls stable for $\kappa\to 1$, albeit much weaker bound than thin-wall Q-balls.
The charge $Q$ is manifestly $q$-independent and actually approaches zero in the thick-wall limit $\omega\to m_\phi$, despite the diverging radius:
\begin{align}
Q \simeq \frac{32\pi \omega}{9\beta^2} \sqrt{m_\phi^2-\omega^2} \, \int_0^\infty \dd x \,x^2 g(x)^2\,,
\end{align}
with the $\beta$ from Eq.~\eqref{eq:phi_potential}.
$p=3$ Q-balls thus become more and more dilute while carrying less and less charge and energy as $\kappa\to 1$, eventually approaching the vacuum solution $\phi =0$.
However, our classical-field description of these $\phi$ bound states eventually breaks down at small $Q$ and needs to be replaced by a quantum-mechanical picture~\cite{Kusenko:1997ad,Graham:2001hr,Postma:2001ea}.

So far we have only shown that $p=3$ Q-balls are stable for small $\kappa$ (thin wall) and near $\kappa=1$ (thick wall). For $q>4$ our analytical descriptions are accurate enough to prove stability, i.e.~$E<m_\phi Q$, over the entire $\kappa$ range. For $q=4$ we have checked numerically that $E<m_\phi Q$ holds for all $\kappa$, as illustrated in Fig.~\ref{Plot:EmQp3q4}.
$p=3$ Q-balls with any integer $q>3$ thus have $E<m_\phi Q$ for any $\kappa\in (0,1)$.

We have restricted our discussion so far to integer $p$ and $q$, for which indeed $p=3<q$ is the only case with stable Q-balls near $\kappa\sim 1$. For non-integer exponents, Refs.~\cite{PaccettiCorreia:2001wtt,Lennon:2021uqu} have shown that Q-balls with $2<p< 10/3$ are stable near $\kappa\sim 1$.
Our results suggest that those Q-balls are actually stable over the entire range $0<\kappa<1$.

\section{Limit of large exponents}
\label{sec:largen}

The case of large exponents, $2\ll p<q$, allows for a qualitatively different approximation than the thin-wall limit above.
As can be seen from Fig.~\ref{fig:potentialpq}, large $p$ and $q$ lead to a very narrow maximum of $V(f)$, positioned at $f\simeq 1$, no matter the value of $\kappa$. The particle then falls down the almost vertical cliff, all the while remaining at $f\simeq 1$, until it reaches the potential minimum. The motion from the minimum to $f=0$ is subsequently described by the easy-to-solve differential equation
\begin{align}
f''(\rho) +\frac{2}{\rho} f'(\rho) - f(\rho) (1-\kappa^2)= 0\,,
\end{align}
where we neglected any $f^p$ or $f^q$ terms since they are highly suppressed in the $f<1$ region.
The large-exponent profile is then simply
\begin{align}
    f_\text{large-$p$} = \begin{cases}
        1\,, & \rho < \tilde R\,,\\
        \frac{\tilde R}{\rho} \exp\left(\sqrt{1-\kappa^2} (\tilde{R}-\rho)\right) , &\rho\geq \tilde R\,,
    \end{cases}
\end{align}
demanding continuity at the point $\rho=\tilde R$ (which is related to the radius by $R\simeq \tilde R + 1/3$ in the stable $\kappa$ regime).
This ansatz is valid for all $\kappa$ in the $2\ll p<q$ limit. To find the remaining $\tilde{R}(\kappa)$ relation, we can use Eq.~\eqref{eq:friction_loss}; notice that the left-hand side of Eq.~\eqref{eq:friction_loss}, $V(f_+)$, is $\kappa^2/2$ for $p,q\to \infty$, just like in the small-$\kappa$ limit. This gives
\begin{align}
    \tilde R = \frac{1+\sqrt{1-\kappa^2}}{\kappa^2}\,,
\end{align}
valid again for all $\kappa$ in the $2\ll p<q$ limit. 
The large-$p$ radius is shown in Figs.~\ref{Plot:RvsKMod} and \ref{Plot:RvsKNPQ}.
The integrals then take the simple forms
\begin{align}
       \int [f'(\rho)]^2\rho^2\dd\rho  &= \frac{\left(\sqrt{1-\kappa ^2}+1\right) \left(\kappa ^2+\sqrt{1-\kappa ^2}+1\right)}{2 \kappa ^4}\,,\nonumber\\
   \int [f(\rho)]^2\rho^2\dd\rho  &= \frac{8-\kappa ^4-4 \kappa ^2+8 \sqrt{1-\kappa ^2}}{6 \kappa ^6 \sqrt{1-\kappa ^2}}\,,
\label{eq:large_p_integrals}
\end{align}
illustrated in Figs.~\ref{Plot:Integrals_n} and \ref{Plot:IntvsKNPQ}.
Solutions to our differential equation in the large exponent limit show a simple and universal behavior. This does not imply that Q-ball energy and charge become independent of $p$ and $q$ in this limit, as the $\phi_0$ and $\omega_0$ in Eq.~\eqref{eq:phi0} depend on the exponents.

The integrals can also be used in Eq.~\eqref{eq:EmQ} to find the stability constraint in the limit of large exponents, shown in Figs.~\ref{Plot:PlotEmQEquidistant} and \ref{Plot:EmQAll}.
This confirms that the critical $\kappa$, satisfying $E=m_\phi Q$, lies in the narrow finite range $(0.8,0.86)$ for all integer $p>3$.

\section{UV completion}
\label{sec:UV}

So far we have worked with the potential $U(|\phi|)$ from Eq.~\eqref{eq:phi_potential}, which contains non-renormalizable terms for even exponents and charge-breaking terms for odd exponents. In this section, we will show how these operators can be obtained in UV-complete models.

We restrict ourselves to the simplest UV completion, consisting of a $U(1)$-charged complex scalar $\phi$ and a real neutral scalar $\psi$, which have the Lagrangian
\begin{align}
\mathcal{L}&=\partial _{\mu } \phi \partial ^{\mu }\phi ^*+\tfrac{1}{2}  \partial _{\mu } \psi \partial ^{\mu }\psi -U[\phi,\psi]
\label{eq.GeneralLagrangian}
\end{align}
with $U(1)$-symmetric scalar potential
\begin{align}
\begin{split}
U[\phi,\psi]&=m_{\phi}^2|\phi|^2+\tfrac{1}{2}m_{\psi}^2\psi ^2+b |\phi| ^4+c |\phi| ^2 \psi \\
&\quad +d |\phi| ^2\psi ^2+e \psi ^3+a \psi ^4.
\end{split}
\label{eq.GeneralPotential}
\end{align}
Here, $m_{\psi}$ and $m_{\phi}$ are the particle masses and $a$, $b$, $c$, $d$, and $e$ are real constants.\footnote{The Wick--Cutkosky~\cite{Wick:1954eu,Cutkosky:1954ru,Nugaev:2016uqd} and Friedberg--Lee--Sirlin~\cite{Friedberg:1976me} models are notable special cases of this Lagrangian that are not covered by our analysis below because they do not have a thin-wall limit in the sense of Coleman -- despite allowing for Q-ball-like solutions.}
The Euler--Lagrange equations for the fields are
\begin{align}
\partial_\mu \partial^\mu \phi +\frac{\partial U}{\partial \phi^*} =0\,, &&
\partial_\mu \partial^\mu \psi +\frac{\partial U}{\partial \psi} =0\,.
\label{eq:eoms}
\end{align}
Once again we are looking for spherically-symmetric localized solutions to these equations, with time dependence $\phi(\vec{x},t)\propto e^{\ii\omega t}$ for $\phi$ and a time-independent $\psi$.
The leading-order thin-wall limit for such a system was recently analyzed in Ref.~\cite{Lennon:2021zzx}, where it was shown that the multi-field generalizations of Eq.~\eqref{eq:minimum} are
\begin{align}
\begin{split}
\left.\left(  \frac{\partial U(|\phi|,\psi)}{\partial |\phi|}
- 2\frac{U(|\phi|,\psi)}{|\phi|}\right) \right|_{|\phi| = \frac{\phi_0}{\sqrt{2}},\,\psi = \psi_0}  &= 0\,,\\
\left.\frac{\partial U(|\phi|,\psi)}{\partial \psi}\right|_{|\phi| = \frac{\phi_0}{\sqrt{2}},\,\psi = \psi_0} &=0\,,
\end{split}
\label{eq.PotentialMinConditions}
\end{align}
which give the Q-ball energy
\begin{align}
E \simeq \omega_0 Q\,, && \text{ with } &&
\omega_0=\sqrt{\frac{U(\phi_0/\sqrt{2},\psi_0)}{(\phi_0/\sqrt{2})^2}}
\label{eq:lennon_E}
\end{align}
to leading order in large $Q$.
The Q-ball properties beyond this thin-wall approximation have to be obtained numerically by solving the coupled non-linear differential equations.
Below, we show that for some special cases of the potential $U[\phi,\psi]$ the two-field system can be mapped onto a one-field system of the form discussed in the previous sections, severely simplifying if not even solving the problem.

Just like in the one-field case explored in the main part of this article, it proves convenient to rescale the fields $\phi$ and $\psi$ by their thin-wall values $\phi_0$ and $\psi_0$ as determined by Eq.~\eqref{eq.PotentialMinConditions}:
\begin{align}
\phi(\vec{x},t)=e^{\ii\omega t}\frac{\phi_0}{\sqrt{2}} f(|\vec{x}|) \,, && \psi(\vec{x},t)=\psi_0 h(|\vec{x}|)\,,
\label{eq.TimeDependenceAnsatz}
\end{align}
where $f(|\vec{x}|)$ and $h(|\vec{x}|)$ are dimensionless functions that are $\leq \mathcal{O}(1)$ for all $\vec{x}$. We furthermore perform the same coordinate transformation as before, $\vec{x}\to \vec{x}\,\sqrt{m_\phi^2 - \omega_0^2}$, with $\omega_0$ from Eq.~\eqref{eq:lennon_E}. $\omega$ is  replaced by $\kappa$ as in Eq.~\eqref{eq:kappa}. All of this ensures that the equation of motion for $f(\rho)$ resembles that of our one-field scenario as closely as possible, except, of course, for the presence of $h(\rho)$.
Since the rescaling is difficult for the general potential $U[\phi,\psi]$ we will only show the resulting differential equations for $f$ and $h$ for some special examples below.

The Q-ball charge $Q$ is determined entirely by the charged field $\phi$ and is again given by our equation~\eqref{eq:charge} upon using the definitions we set forth.
The Q-ball energy, on the other hand, contains a contribution from the neutral field $\psi$:
\begin{align}
E &= \int \dd^3 x \left[\frac{\phi_0^2 (f')^2}{2} + \frac{\psi_0^2 (h')^2}{2}  +\frac{ \omega^2\phi_0^2f^2}{2} + U[\phi,\psi]\right]\nonumber\\
&= \omega Q +\frac{4\pi}{3\sqrt{m_\phi^2-\omega_0^2}} \int \dd \rho\, \rho^2 \left[\phi_0^2 (f')^2 +  \psi_0^2 (h')^2\right] ,    
\label{eq:two-field_energy}
\end{align}
where in the second line we used the virial theorem, e.g.~Refs.~\cite{Derrick:1964ww,Friedberg:1976me}.

\subsection{Massive \texorpdfstring{$\psi$}{psi}}
\label{sec:massive}

Assuming $m_{\psi} \gg m_{\phi}$ we can neglect the kinetic term in the Euler--Lagrangian equation associated with the field $\psi$. Thus we end up with $\partial U[\phi,\psi]/ \partial \psi=0$. We can solve the latter order-by-order in large $m_\psi$ with the following ansatz 
\begin{align}
\psi = \frac{x_1}{m_\psi^2}+\frac{x_2}{m_\psi^4}+\frac{x_3}{m_\psi^6}+\dots
\label{eq.Ansatz}
\end{align}
with coefficients $x_j$ that depend on $|\phi|^2$ and the coefficients in the two-field potential. The $\psi$ field is hence suppressed compared to the $\phi$ field in this expansion, which suppresses $\psi$'s contribution to the Q-ball energy.
After solving $\psi$'s equation of motion and plugging the resulting $\psi$ back into the potential we get the potential for $\phi$:
\begin{align}
U[\phi] &=m_\phi^2 |\phi| ^2+ \left(b-\frac{c^2}{2 m_\psi^2}\right)|\phi| ^4+ \left(\frac{c^2
   d}{m_\psi^4}-\frac{c^3 e}{m_\psi^6}\right)|\phi| ^6 \nonumber\\
&\quad+ \left(\frac{c^4 a+6 c^3 d e}{m_\psi^8}-\frac{2 c^2
   d^2}{m_\psi^6}\right)|\phi| ^8+\frac{4 c^2 d^3 }{m_\psi^8}|\phi| ^{10} \nonumber\\
   &\quad+\mathcal{O}\left(\frac{1}{m_\psi^{10}}\right) .\label{eq.PotentialUpto8}
\end{align}
As expected for an effective field theory at tree level, we find higher-dimensional operators in $|\phi|^2$ suppressed by powers of $m_\psi^2$. 
Below we show some examples that can be approximately described by our one-field potential from Eq.~\eqref{eq:phi_potential}.
An alternative derivation of the same cases that highlights the proper expansion parameter is deferred to App.~\ref{app:rescaled_heavy} for the curious reader.

\subsubsection{Large \texorpdfstring{$m_\psi$, $a=e=0$}{mpsi=0, a=e=0}}
\label{sec:heavy46case}

Setting $a=e=0$ and only keeping the terms up to $\mathcal{O}\left(1/m_\psi^{4}\right)$, we find
\begin{align}
&U[\phi]\simeq m_{\phi }^2|\phi| ^2 + \left(b-\frac{c^2}{2 m_\psi ^2}\right)|\phi|^4 +\frac{c^2 d }{m_{\psi }^4}|\phi| ^6.
\label{eq:approx46}
\end{align}
This corresponds to the $p=4$, $q=6$ case of Eq.~\eqref{eq:phi_potential} with coefficients $\beta = -b+\frac{c^2}{2 m_\psi ^2}$ and $\xi=\frac{c^2 d }{m_{\psi }^4}$.
Since $\beta$ is required to be positive, we have to assume that $b$ is of order $\mathcal{O}\left(1/m_\psi^{2}\right)$. Both $\beta$ and $\xi$ are hence suppressed in this expansion, with $\xi$ being of order $\beta^2$. Eq.~\eqref{eq:phi0} shows that $ m_\phi^2-\omega_0^2 = \beta^2/(4\xi)$ is of order $\mathcal{O}\left(m_\psi^{0}\right)$ so $\omega_0$ can naturally take any value between $0$ and $m_\phi$.

\subsubsection{Large \texorpdfstring{$m_\psi$, $e=d=0$}{mpsi=0, e=d=0}}

Setting $d=e=0$ and keeping terms up to $\mathcal{O}\left(1/m_\psi^{8}\right)$ we find the potential
\begin{align}
&U[\phi]\simeq m_{\phi }^2|\phi| ^2 + \left(b-\frac{c^2}{2 m_\psi ^2}\right)|\phi|^4 +\frac{c^4 a }{m_{\psi }^8}|\phi| ^8,
\end{align}
which corresponds to Eq.~\eqref{eq:phi_potential} with $p=4$ \& $q=8$ and $\beta=-b+\frac{c^2}{2 m_\psi ^2}$ and $\xi=\frac{c^4 a }{m_{\psi }^8}$. Once again both $b$ and $\beta$ are $\mathcal{O}\left(1/m_\psi^{2}\right)$ in the heavy-$\psi$ expansion. In this case, we find that $ m_\phi^2-\omega_0^2 \propto \beta^{3/2}/\sqrt{\xi}$ is of order $\mathcal{O}\left(m_\psi\right)$ and hence large. Generically we then expect Q-balls with $\omega_0\ll m_\phi$.

\subsubsection{Large \texorpdfstring{$m_\psi$, $d=0$, $b=c^2/(2m_\phi)$}{mpsi=0, d=0, b=c c/2mphi}}

Setting $d=0$ and $b=c^2/2 m_\psi^2$ yields
\begin{align}
&U[\phi]\simeq m_{\phi }^2|\phi| ^2 -\frac{c^3 e }{m_{\psi }^6}|\phi|^6+\frac{c^4 a }{m_{\psi }^8}|\phi| ^8,
\end{align}
up to $\mathcal{O}\left(1/m_\psi^{8}\right)$ terms.
This gives $p=6$, $q=8$ with parameters $\beta=\frac{c^3 e }{m_{\psi }^6}$ and $\xi=\frac{c^4 a }{m_{\psi }^8}$.
Here, $ m_\phi^2-\omega_0^2 \propto \beta^{3}/\xi^2$ is of order $\mathcal{O}\left(1/m_\psi^2\right)$ and hence small, so $\omega_0$ should be of order $m_\phi$.

\subsection{Massless \texorpdfstring{$\psi$}{psi}}
\label{sec:massless}

The heavy-$\psi$ framework from above unsurprisingly generates even exponents $p$ and $q$. Considering instead a \emph{massless} $\psi$ can give odd or even rational exponents as we will show below. For these scenarios we work with the equations of motion from Eq.~\eqref{eq:eoms} rather than the potential. We go through two simple cases below.

\subsubsection{\texorpdfstring{$m_\psi=d=e=0$}{mpsi=d=e=0}}

Performing the above-mentioned rescaling for the case $m_\psi=d=e=0$ yields the following simple equations of motion for $f(\rho)$ and $h(\rho)$
\begin{align}
&h''(\rho)+\frac{2 h'(\rho )}{\rho } + 2 \sqrt{\frac{a}{b}} \left[f(\rho )^2-h(\rho )^3\right]=0\,,\label{eq.EOMmde0}\\
&f''(\rho )+\frac{2 f'(\rho )}{\rho }+f(\rho )\left(\kappa ^2-1+2 h(\rho )\right) -f(\rho )^3 =0\,,
\nonumber
\end{align}
which depend only on two parameters: $\kappa$ and $a/b$. 
If we choose $a\gg b$ we can neglect the derivatives in the $h$ equation and solve the equation of motion algebraically to $h(\rho )=f(\rho )^{2/3}$. Plugging this into the second equation we recover for $f(\rho)$ exactly the single-field equation~\eqref{eq:eom} with $p=8/3$ and $q=4$.

\begin{figure}[tb]
	\includegraphics[width=0.48\textwidth]{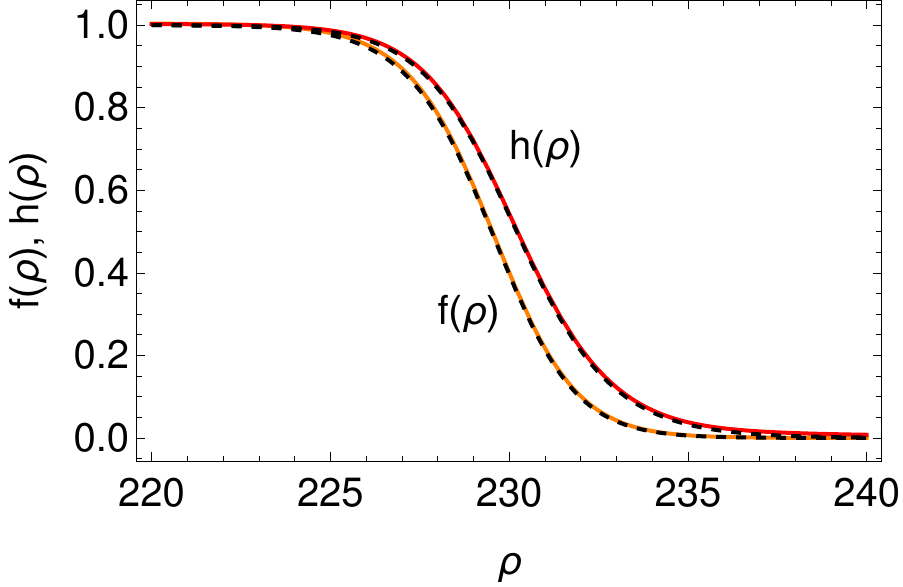}
	\caption{ Profiles for $f(\rho)$ and $h(\rho)$ solving Eq.~\eqref{eq.EOMmde0} numerically for $\kappa = 0.05$ and $2 \sqrt{a/b} = 1000$. The dashed black lines show our thin-wall predictions, i.e.~$f(\rho)$ from Eq.~\eqref{eq:profile2Mod} with $n(8/3,4)\simeq 0.8$ and $h(\rho) = f(\rho)^{2/3}$.
}
	\label{Plot:two-field_profile}
\end{figure}

The relation $h(\rho )=f(\rho )^{2/3}$ holds for all $\kappa$, allowing us to solve the two-field system by simply solving the single-field equation~\eqref{eq:eom}. Of course, for small or large $\kappa$ we can even approximate  $f(\rho)$ and hence $h(\rho)$ using our analytical results from above.
As illustrated in Fig.~\ref{Plot:two-field_profile}, the two profiles (and radii) are indeed very well described by our transition profile from Eq.~\eqref{eq:profile2Mod} using $n(8/3,4)\simeq 0.8$ from Eq.~\eqref{eq:NPQ}, at least for small $\kappa$ and large $a/b$.
This also allows us to obtain analytic approximations for Q-ball energy and charge.
Notice that $\psi_0/\phi_0 \propto (b/a)^{1/4}$ here, so $\psi$'s contribution to the Q-ball energy~\eqref{eq:two-field_energy} is suppressed in the limit of interest.\footnote{Although it is not difficult to keep the contribution; in the thin-wall limit, the relevant integral for $h(\rho)=f(\rho)^k$ is
$\int\dd\rho \,\rho^2 \left(h'\right)^2\simeq  n k R^2/(2n+4k)$.}
Since these Q-balls are then approximately single-field Q-balls with $p=8/3 < 10/3$, they have stable thin- and thick-wall limits~\cite{PaccettiCorreia:2001wtt,Lennon:2021uqu} and are hence stable for all $\kappa$, allowing for arbitrarily large or small charge $Q$. 
This case therefore provides one of the simplest renormalizable realizations of a Q-ball that can grow naturally via accumulation of particles without requiring a minimal threshold charge.
Of course, for small $Q$ our classical analysis needs to be replaced by a quantum one.

\subsubsection{\texorpdfstring{$m_\psi=d=a=0$}{mpsi=d=a=0}}

Next, let us consider $m_\psi=d=a=0$. This gives the system of equations
\begin{align}
&h''(\rho)+\frac{2 h'(\rho )}{\rho } + \frac{9 e}{c} \left[f(\rho )^2-h(\rho )^2\right]=0\,,\label{eq.EOMmda0}\\
&f''(\rho )+\frac{2 f'(\rho )}{\rho }+f(\rho ) \left(-2 f(\rho )^2+\kappa ^2+3 h(\rho )-1\right) =0\,.
\nonumber
\end{align}
Now, by choosing $e\gg c$ we see from the first equation that the two profiles will approximately coincide: $h(\rho )=f(\rho )$. After plugging this into the second equation we recover Kusenko's single-field case with $p=3$ and $q=4$ for $f(\rho)$~\cite{Kusenko:1997ad}.
The field ratio $\psi_0/\phi_0\propto\sqrt{|c/e|}$ is again suppressed in the limit of interest.
Just like in the previous case, we hence find a simple renormalizable realization of a stable Q-ball with arbitrary charge.

\section{Discussion and Conclusion}
\label{sec:conclusion}

Q-balls are simple examples of bound states consisting of scalars $\phi$. Assuming an attractive self-interaction in the scalar potential, these objects can contain a large number of particles, allowing for a classical description. Q-balls have been conceived many decades ago, but their description outside of the simplest of limits has proven challenging, owing to the non-linear nature of their underlying field equation.
In this article, we performed an exhaustive study of Q-balls generated by three-term potentials of the form
\begin{align}
U(|\phi|) = m^2_\phi |\phi|^2 -\beta |\phi|^p+\xi |\phi|^q \,.\nonumber
\end{align}
For $2<p<q$ and positive $\beta$ and $\xi$, these are the simplest potentials that can give large Q-balls {\`a} la Coleman. We have provided analytical approximations that describe stable Q-balls for \emph{all} exponents $p$ and $q$, in part by generalizing the procedure of Ref.~\cite{Heeck:2020bau}.

We find a surprisingly universal Q-ball behavior that depends only weakly on the integers $p$ and $q$:
i) The instability threshold where $E=m_\phi Q$ falls in the narrow range $\kappa\in (0.80,0.86)$ for all $p>3$.
ii) The volume energy does not depend on $p$ and $q$, and even the surface energy shows only a mild dependence.
iii) Radii of stable Q-balls with $p>3$ scale with $1/\kappa^2$ up to an $\mathcal{O}(1)$ prefactor that depends on $p$ and $q$. Furthermore, all stable Q-balls  have radii $R>1$, or, in terms of the actual dimensionful Q-ball radius,
\begin{align}
    R_\text{Q-ball} > 1/\sqrt{m_\phi^2 - \omega_0^2}\,.
\end{align}
In particular, $ R_\text{Q-ball}>1/m_\phi$, in perfect agreement with the bound state conjecture of Ref.~\cite{Freivogel:2019mtr} for the radius of any stable bound state. 

The discussion of single-field Q-balls is unavoidable an effective one, as there are no values for $p$ and $q$ that lead to a renormalizable charge-conserving potential that is bounded from below. To highlight that our analysis is nevertheless useful, we studied a simple renormalizable two-field model that can be effectively described by our one-field scenario with several $p$ and $q$, including -- quite surprisingly -- odd and fractional exponents. Repeating this analysis for models with more fields  would undoubtedly allow us to generate potentials with an even wider range of exponents.

Finally, our results for the ground-state profiles of global Q-balls can be generalized to excited states~\cite{Almumin:2021gax} as well as gauged and Proca Q-balls via the mapping relations of Refs.~\cite{Heeck:2021zvk,Heeck:2021bce}.

\section*{Acknowledgements}
We thank Chris Verhaaren and Arvind Rajaraman for comments on the manuscript.
This work was supported in part by the National Science Foundation under Grant PHY-2210428.

\appendix

\section{Alternative derivation of the heavy \texorpdfstring{$\psi$}{psi} cases}
\label{app:rescaled_heavy}

If the expansion in large $m_\psi$ in Sec.~\ref{sec:massive} did not seem convincing, we will provide an alternative derivation here that follows the procedure of Sec.~\ref{sec:massless} by first rescaling the two fields.

We start with the case $a=e=0$. It proves convenient to replace $c = \sqrt{2}m_\psi \sqrt{b+\beta}$, with $\beta$ defined just as below Eq.~\eqref{eq:approx46}. All the rescaling can be performed exactly, but since the limit of interest will be small $\beta$ we only show the equations in that limit here. Eq.~\eqref{eq.PotentialMinConditions} can be solved to give
\begin{align}
\phi_0 \simeq \frac{m_\psi }{\sqrt{2}\sqrt{b d}} \sqrt{\beta}\,, &&
\psi_0 \simeq \frac{m_\psi }{2\sqrt{2 b} d} \beta\,.
\end{align}
In particular, $\psi$ is suppressed compared to $\phi$ by $\sqrt{\beta/d}$, so $\psi$'s contribution to the Q-ball energy will be small.
To leading order in small $\beta$, the equation of motion for $h(\rho)$ takes the form
\begin{align}
&h''(\rho)+\frac{2 h'(\rho )}{\rho } + \frac{8 b d }{\beta^2} \left[h(\rho )-f(\rho )^2\right] =0\,
\end{align}
and thus fixes $h(\rho) = f(\rho)^2$ as long as $\beta^2\ll b d$.
The equation of motion for $f$ with $h(\rho) = f(\rho)^2$ then matches the single-field Eq.~\eqref{eq:eom} with $p=4$ and $p=6$, plus terms that are suppressed by $\beta/b$.
This matches the conclusion of Sec.~\ref{sec:heavy46case} but highlights that the expansion parameter is not really large $m_\psi$  but rather small $\beta$. Of course, we have identified $\beta$ as being of order $m_\psi^{-2}$ above, so this is consistent.

The discussion of the case $d=e=0$ is analogous. We again replace $c$ by $\beta$ and go to the small $\beta$ limit, which gives $\psi_0/\phi_0\propto (\beta/a)^{1/4}$, so $\psi$ is again suppressed compared to $\phi$. For small $\beta$, the equation of motion for $\psi$ gives $h(\rho)=f(\rho)^2$. The differential equation for $f(\rho)$ matches our single-field equation with $p=4$, $q=8$ up to terms suppressed by $\beta/b$.

Finally, the case with $b=c^2/(2m_\psi^2)$, $d=0$ is slightly more laborious but analogous. We expand in small $e$, which is equivalent to small $\beta$. The field ratio $\psi_0/\phi_0\propto \sqrt{e c/a}/m_\psi$ is suppressed again, and again we find $h(\rho)=f(\rho)^2$ for small $e$. The differential equation for $f(\rho)$ matches the $p=6$, $q=8$ case plus terms suppressed by $e^2/(a m_\psi^2)$.

\bibliographystyle{utcaps_mod}
\bibliography{BIB}

\end{document}